\documentclass{vietnam}

\def\be{\begin{equation}}
\def\ee{\end{equation}}
\def\bea{\begin{eqnarray}}
\def\eea{\end{eqnarray}}

\begin{document}


\title{Magnetic Ribbons: A Minimum Hypothesis Model for Filaments}

\author{Sayantan Auddy, Shantanu Basu, and Takahiro Kudoh}

\address{Department of Physics and Astronomy, The University of Western Ontario, London, Canada.}

\maketitle
\begin{figure}
\vskip -0.5cm
\centering
$\begin{array}{cc}
\includegraphics[angle=0,height=8.cm]{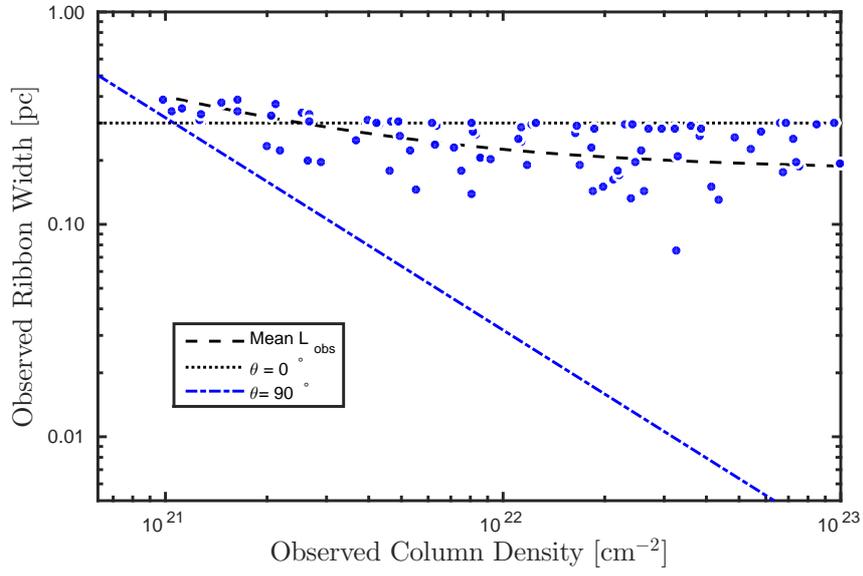} 
\end{array}$
\caption{Apparent ribbon width $L_{\rm obs}$ versus observed column density $N_{\rm obs}$. Each blue dot corresponds to a magnetic ribbon with intrinsic column density $N$ and observing angle $\theta$, chosen randomly  . The black dashed line is the mean ribbon width for the entire range of values of $N_{\rm obs}$. The black dotted line is the width when the ribbon is viewed at $ \theta = 0^{\circ}$. The blue dot-dashed line is the width for the side on view i.e., $\theta = 90^{\circ}$, where $L_{\rm{obs}} = 2c^{2}_{s} / (G \Sigma)$ is essentially the Jeans length.}
\label{fig:clm}
\end{figure}

\abstracts{
We develop a magnetic ribbon model for molecular cloud filaments. These result from turbulent compression in a molecular cloud in which the background magnetic field sets a preferred direction. We use our model to calculate a synthetic observed relation between apparent width in projection versus observed column density. The relationship is relatively flat, in rough agreement with the observations, and unlike the simple expectation based on a Jeans length argument.
}
\textit{Herschel} has revealed many filamentary structures in molecular clouds \cite{and10}, many of which have a magnetic field aligned perpendicular to the long axis \cite{pal13}. Much of the theoretcal interpretation of these filaments is as isothermal cylinders which are either in equilibrium or unstable to collapse depending on their mass per unit length. However this is challenged by the existence of a large scale magnetic field perpendicular to the long axis unless the field strength is dynamically insignificant.  Further observational results \cite{arz11} found that the width versus the column density of the filaments was a flat relation, unlike a simple expectation based on a Jeans length. This naturally motivates us to interpret a filament as a magnetic ribbon, a triaxial object that is flattened along the direction of the large scale magnetic field. 
These result from turbulent compression in a molecular cloud with a subcritical mass-to-flux ratio, in which the background magnetic field sets a preferred direction. 
We consider local pressure balance between the magnetic pressure and the ram pressure of the flow in a compressed region of a subcritical cloud \cite{aud16}. The magnetic field $B$ is aligned perpendicular to the long axis ($y$) of the massive star forming ribbon and this leads to flattening along the $z$-direction parallel to $B$. The cloud is stratified in the $z$-direction, with turbulent compression happening primarily in  the $x$-direction. The width of a ribbon turns out to be independent of the density of the medium and depends only on the initial compression scale and Alfv{\'e}nic Mach number of the turbulence \cite{aud16}. The thickness along the magnetic field direction is essentially the Jeans length and does depend on the density. 
Due to the triaxial symmetry, the observed width of the ribbon depends on the viewing angle \cite{aud16}. Observed from a random set of viewing angles, the relation between the observed ribbon width and observed column density is relatively flat as shown in Figure. \ref{fig:clm} (unlike expectations based on the Jeans length) over the range $10^{21}\, {\rm cm}^{-2} - 10^{23}\, {\rm cm}^{-2}$, in rough agreement with the observations \cite{arz11}.
%

We have presented a simple model in which a magnetic ribbon has a thickness that is set by the standoff between ram pressure and the magnetic pressure. Gravitationally driven ambipolar diffusion then leads to runaway collapse of the densest regions in the ribbon, where the mass-to-flux ratio has become supercritical.




\end{document}